\documentclass[rnaas]{aastex63}

\newcommand{\rf}{\emph{realfast}}
\newcommand{\dmunits}{{\rm pc\,cm^{-3}}}

\newcommand{\src}{FRB\,180916}
\usepackage{gensymb}

\submitjournal{RNAAS}
\shorttitle{Realfast detection of \src}
\shortauthors{Aggarwal et al.}
\graphicspath{{./}{figures/}}

\begin{document}

\title{VLA/realfast detection of burst from FRB180916.J0158+65 and Tests for Periodic Activity}

\correspondingauthor{Kshitij Aggarwal}
\email{ka0064@mix.wvu.edu}

\author[0000-0002-2059-0525]{Kshitij Aggarwal}
\affiliation{Center for Gravitational Waves and Cosmology, West Virginia University, Chestnut Ridge Research Building, Morgantown, WV 26505}
\affiliation{Department of Physics and Astronomy, West Virginia University, Morgantown, WV 26506}

\author[0000-0002-4119-9963]{Casey J. Law}
\affiliation{Cahill Center for Astronomy and Astrophysics, MC 249-17 California Institute of Technology, Pasadena, CA 91125, USA}

\author[0000-0003-4052-7838]{Sarah Burke-Spolaor}
\affiliation{Center for Gravitational Waves and Cosmology, West Virginia University, Chestnut Ridge Research Building, Morgantown, WV 26505}
\affiliation{Department of Physics and Astronomy, West Virginia University, Morgantown, WV 26506}
\affiliation{CIFAR Azrieli Global Scholars program, CIFAR, Toronto, Canada}

\author[0000-0003-4056-9982]{Geoffrey Bower}
\affiliation{ASIAA, 645 N. A'ohoku Pl, Hilo, HI 96720}
\affiliation{Affiliate Graduate Faculty, UH Manoa Physics and Astronomy}

\author[0000-0002-5344-820X]{Bryan J. Butler}
\affiliation{National Radio Astronomy Observatory}

\author[0000-0002-6664-965X]{Paul Demorest}
\affiliation{National Radio Astronomy Observatory}

\author[0000-0002-3873-5497]{Justin Linford}
\affiliation{National Radio Astronomy Observatory, Socorro, NM, 87801, USA}

\author{T. J. W. Lazio}
\affiliation{Jet Propulsion Laboratory, California Institute of Technology, M/S 67-201, 4800 Oak Grove Dr., Pasadena, CA  91109}

\begin{abstract}
We report on the detection of a burst from \src\ by \rf/VLA and present software for interpreting FRB periodicity. We demonstrate a range of periodicity analyses with bursts from \src, FRB\,121102 and FRB\,180814. Our results for \src\ and FRB\,121102 are consistent with published results. For FRB\,180814, we did not detect any significant periodic episodes. The \rf-detected and other high-frequency bursts for \src\ tend to lie at the beginning of the activity window, indicating a possible phase-frequency relation. The python package \texttt{frbpa} can be used to reproduce and expand on this analysis to test models for repeating FRBs.
\end{abstract}


\section{Realfast Detection of \src} \label{sec:intro}
\defcitealias{pr3}{PR3}
\defcitealias{rfr3atel}{rfATel}
One of the repeating fast radio bursts (FRBs), \src, has been found to have periodic episodes of higher activity, with a period of 16.35 days (\citealt{pr3}, hereafter PR3).
Its bursts are clustered in a 4-day phase window with some cycles showing no bursts, while others show multiple bursts. 


On April 23 at 20:11 UTC, we used the \rf\ commensal fast transient search system at the Karl G. Jansky Very Large Array (VLA) to observe \src. We detected a burst with a S/N of 13 at a DM of 349.8$~\dmunits$ (\citealt{rfr3atel}, hereafter rfATel). The \rf\ system localised the burst in realtime to a location of J2000 R.A.$=01^{\rm h}58{^{\rm m}}00{^{\rm s}}.634$, Decl.=$-65\degree 43'00''.6331$. The position of this burst is consistent with the reported localization by EVN \citep{marcote2020}, given the VLA localization precision of 0.8$\arcsec$. The details of the burst were reported in \citetalias{rfr3atel} .

\section{Periodicity Analysis Techniques}
Since the discovery of periodicity in the activity of \src, many observatories have reported detections of bursts from this FRB \citep{chawla2020, marcote2020, Pilia2020, scholz2020, gmrt2020}. In total, 19 bursts have been detected at telescopes other than CHIME from \src. This, along with the bursts reported by \citetalias{pr3}, leads to a sample of 51 bursts that have been used in the analysis reported here.

Following the procedure in \citetalias{pr3} and considering a period of 16.35 days, we generated the phase histogram of all published bursts from \src\ (Fig.~\ref{phase_fig}, top panel). Most of the bursts lie within a 4 day (or 4/16 = 0.25 phase) phase window from phase 0.4 to 0.6.
Our VLA detection lies at a phase of 0.3, which is the earliest phase at which a burst has been detected so far. The addition of bursts from telescopes other than CHIME makes the phase distribution more symmetric. We also generated the phase histograms at other periods within the period error reported by \citetalias{pr3}, which resulted in a similar conclusion. 

We used three tests to search for episode periodicity in this burst sample. First, we used the Pearson Chi-square test done by \citetalias{pr3}. Second, we followed the approach of \citet{Rajwade2020} to search for the period with a folded profile of minimum fractional width. We also use the Quadratic-Mutual-Information-based periodicity search technique \citep{Huijse2018} implemented in P4J\footnote{\url{https://github.com/phuijse/P4J}} to search for a period in these bursts. All three search techniques were used on all 51 bursts, and also on 32 CHIME bursts. Following \citetalias{pr3}, we also searched for a periodicity after binning the data to obtain just the ``activity days". 

All the scripts developed for periodicity search and phase analysis reported here are available as a python package \texttt{frbpa}\footnote{\url{https://github.com/KshitijAggarwal/frbpa}}. \texttt{frbpa} has various functions that can be used to search for periodicity in the activity of repeating FRBs. It can also be used to visualize the dependence of the burst MJDs and observations on phases.

\section{Results and Discussion}
For all the methods listed above, we recovered a period that was consistent with the results of \citetalias{pr3}, using all the bursts and using just the ``activity days". 
We also extended the analysis of aliasing in \citetalias{pr3} to include bursts from all other telescopes. As the periodicity at all the frequencies is expected to be same, the standard deviation of the burst phases should be minimum at the optimal period (or its alias), which was found to be the case at the published period\footnote{We also followed these methods after including the openly available but unpublished CHIME burst sample from \url{https://www.chime-frb.ca}, which lead to similar conclusions}.

We also used \texttt{frbpa} to search for periodicity in two other active repeaters: FRB\,121102 (R1), and  FRB\,180814 (R2). \citet{Rajwade2020} reported a period of 157 days in R1 using a sample of 235 bursts detected over a time span of 7 years. We used P4J to search for periodicity on this sample and recovered a period consistent with their results. For R2, we used 21 bursts detected by CHIME\footnote{\url{https://www.chime-frb.ca}}.
We did not detect any significant period, using all the three techniques for periodicity search on these bursts. Moderately significant detections at activity windows of 33, 45, 90 and 138 days were observed. 
Further burst detections are needed to verify the periods reported for this FRB.

\begin{figure}
\label{phase_fig}
\includegraphics[scale=0.7]{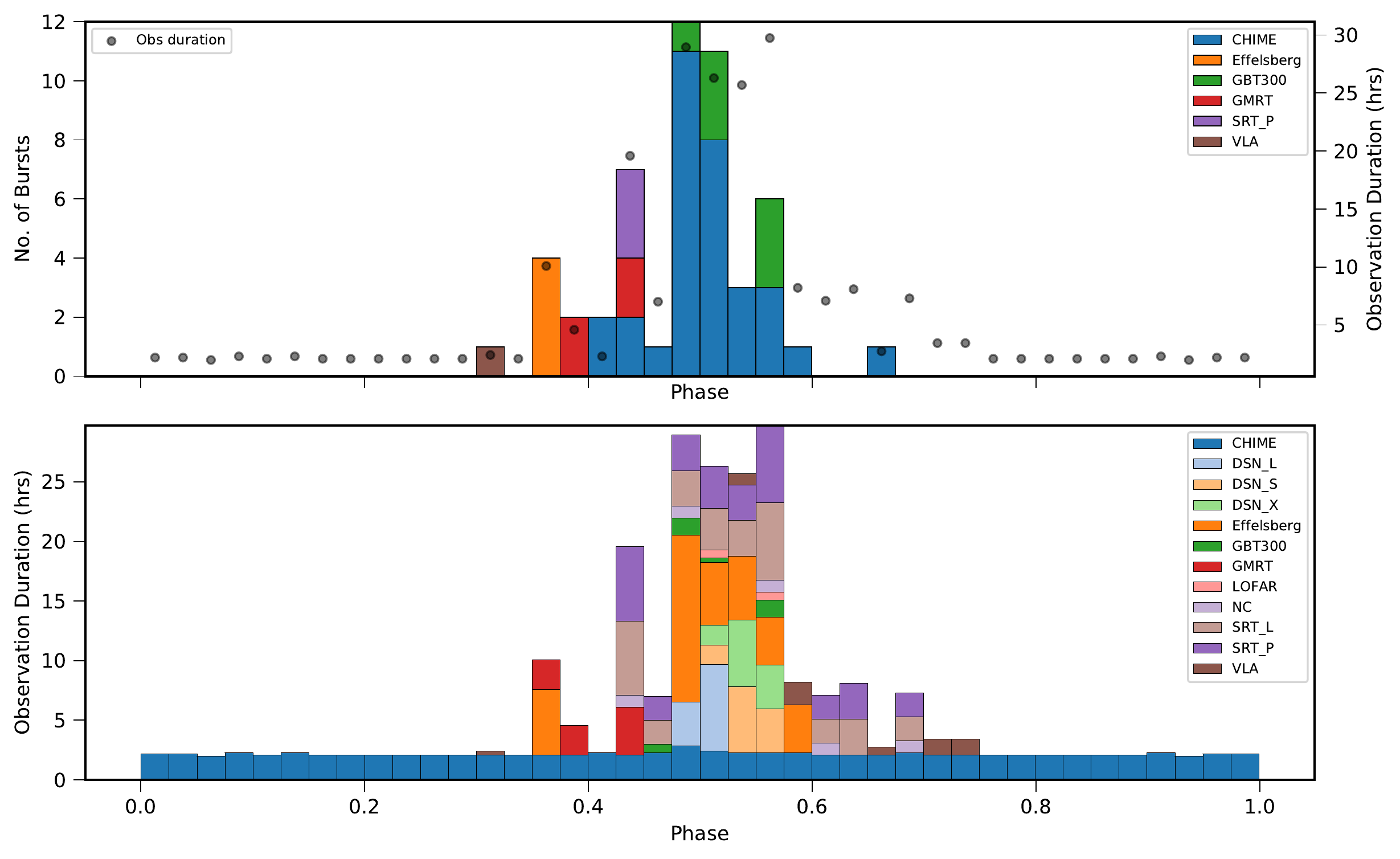}
\caption{Burst detection and observation histograms for \src, assuming a period of 16.35 days. Different colors represent different observatories. \textit{Top:} Stacked histogram of the detected bursts relative to phases for all published \src\ bursts. The black dots show the total observation duration for each phase, summed for all the telescopes in the bottom panel. \textit{Bottom:} Stacked histogram of observation duration with respect to phase. 
}
\end{figure}

We also note that for \src, the bursts detected at high frequency (i.e above 600MHz; detected using VLA and Effelsberg) are at a lower phase value than most of the low-frequency bursts (Fig.~\ref{phase_fig}, top panel). This indicates that there may be a correlation between frequency and phase, where high-frequency emission is suppressed at higher phase values. 
Although these high-frequency observations have good coverage within the activity phases (Fig.~\ref{phase_fig}, bottom panel) with good sensitivity, the detections have only occurred at phase $<$0.4. This cannot be explained by the models which invoke an interacting neutron-star-binary system to explain the periodicity. These models predict a wider activity window at higher frequencies, as high-frequency photons are generally transmittable \citep{ioka2020}. 

Therefore, more high-frequency observations across different activity phases would be imperative to comment on the periodicity at high frequency. Moreover, with the detection of more bursts from \src\ and other repeaters in the future, more confident analysis of plausible periodicity (and phase-dependent burst rates) in their activity would be possible. 

\section{Acknowledgement}
SBS and KA acknowledge support by NSF grant 1714897. CJL acknowledges support under NSF grant 2022546. The National Radio Astronomy Observatory is a facility of the National Science Foundation operated under cooperative agreement by Associated Universities, Inc. We acknowledge use of the CHIME/FRB Public Database, provided at https://www.chime-frb.ca/ by the CHIME/FRB Collaboration.

\vspace{5mm}
\facilities{EVLA}

\software{rfpipe \citep{2017ascl.soft10002L}, 
fetch \citep{agarwal2019}, 
numpy \citep{numpy}, matplotlib \citep{Hunter:2007}
}

\bibliography{r3}{}
\bibliographystyle{aasjournal}



\end{document}